\documentstyle[epsf]{l-aa}

\begin{document}

\thesaurus{11.01.2; 11.02.2; 11.16.1; 13.07.3; 13.21.1}

\title{Optical--IUE observations of the gamma--ray loud
BL Lacertae object S5 0716+714: data and interpretation}

\author{G. Ghisellini\inst{1} \and M. Villata\inst{2} \and
C.M. Raiteri\inst{2} \and S. Bosio\inst{3} \and
G. De Francesco\inst{2} \and G. Latini\inst{3} \and
M. Maesano\inst{4} \and E. Massaro\inst{4} \and 
F. Montagni\inst{4} \and R. Nesci\inst{4} \and
G. Tosti\inst{5} \and M. Fiorucci\inst{5} \and
E. Pian\inst{6} \and L. Maraschi\inst{7} \and
A. Treves\inst{8} \and A. Comastri\inst{9} \and
M. Mignoli\inst{9}}

\offprints{G. Ghisellini}

\institute{
Osservatorio Astronomico di Brera, via Bianchi, 46, 22055 Merate, Italy
\and
Osservatorio Astronomico di Torino, 10025 Pino Torinese, Italy
\and
Universit\`a di Torino, via Giuria 1, Torino, Italy
\and
Istituto Astronomico, Universit\`a di Roma, Unit\`a GIFCO--CNR Roma I,
via Lancisi, 29, Roma, Italy
\and
Osservatorio Astronomico, Universit\`a di Perugia, Perugia, Italy
\and
Space Telescope Science Institute, Baltimore, USA
\and
Osservatorio di Brera, via Brera, 28, Milano, Italy
\and
International School for Advanced Studies, Trieste, Italy
\and
Osservatorio Astronomico di Bologna, Italy 
}

\date{Received date: 31 October 1996; accepted date: 3 June 1997}

\maketitle
 
\markboth{Ghisellini et al.: Optical-IUE observations of S5 0716+714...}{...}

\begin{abstract}
We monitored the BL Lac object S5 0716+714 in the optical band during
the period November 1994 -- April 1995, which includes the time of 
a $\gamma$--ray observation by the Energetic Gamma Ray Experiment 
Telescope (EGRET) on February 14--28, 1995. 
The light curves in the $R$ and $B$ bands show fast fluctuations
superimposed on longer timescale variations. 
The color index correlates with intensity during the rapid flares
(the spectrum is flatter when the flux is higher), but it is rather 
insensitive to the long term trends. 
Over the 5 month observational period the light curve shows  
an overall brightening of about 1 mag followed by a fast decline. 
The EGRET pointing covers part of the very bright phase ($V\sim 13.2$)
and the initial decline. 
An ultraviolet spectrum was also obtained with the International Ultraviolet 
Explorer (IUE) (1200--3000 \AA) during the EGRET observations. 
The variability of the optical emission  of S5 0716+714 by itself
sets important constraints on the magnetic field strength and on the physical 
processes responsible for it. 
Interpreting the whole electromagnetic spectrum with synchrotron self
Compton models leads to the prediction of a bright $\gamma$--ray state 
during the EGRET pointing. 
We discuss how the $\gamma$--ray data could be used as a diagnostic
of the proposed models. 


\end{abstract}

\section{Introduction}
One of the most important recent discoveries in the field of active
galactic nuclei (AGN) is that flat spectrum radio sources (blazars)
emit a substantial fraction, sometimes most of their power, 
at $\gamma$--ray energies. 
About 60 flat  spectrum radio sources have been detected up to now
above 100 MeV by the Energetic Gamma Ray Experiment Telescope (EGRET) 
onboard the Compton Gamma Ray Observatory (CGRO) 
(Fichtel 1994; von Montigny et al. 1995; Thompson et al. 1995).
Strong and rapid variability seems to be a common property of these sources
in the $\gamma$--ray band (e.g. von Montigny et al. 1995). 
The origin of this high energy emission is still unclear.
A point of agreement is that the emitting plasma is in relativistic
motion at small angles to the line of sight, with consequent beaming of 
the emitted radiation, which is required to avoid strong suppression of 
$\gamma$--rays due to photon--photon absorption (McBreen, 1979;
for application to blazars see, e.g.,
Maraschi, Ghisellini \& Celotti 1992; Dondi \& Ghi\-sellini 1995).

The correlated optical and $\gamma$--ray variability observed
in some objects (Wagner et al. 1995, Hartman et al. 1996)
indicates that the same population 
of electrons could be responsible for the flux in both bands, and 
suggests that the optical emission is due to the synchrotron process, 
while the $\gamma$--rays are produced by the inverse Compton process 
between highly relativistic electrons and soft seed photons.
However, the origin of the seed photons, the location and size of the 
emitting region(s), and the degree of relativistic beaming of the high 
energy radiation are unknown.
According to one interpretation, the seed photons can be produced by 
synchrotron emission internally to the emitting blob or jet (Maraschi, 
Ghisellini \& Celotti 1992). 
Another possibility is that the blob itself could illuminate a portion of 
the broad line clouds, whose reprocessed line radiation can dominate the 
radiation energy density in the blob (Ghisellini \& Madau 1996). 
In another scenario the seed photons are produced externally, by
the accretion disk (Dermer \& Schlickeiser 1993), by the broad line region 
illuminated by the disk (Sikora, Begelman \& Rees 1994), and/or by some 
scattering material surrounding the jet (Blandford 1993, Blandford
\& Levinson 1995).
Finally, a dusty torus 
could provide IR radiation
to be scattered in the $\gamma$--ray band, as suggested by Wagner et al. 
(1995).

For plausible values of the parameters, the relativistic electrons which
emit $\gamma$--rays by the inverse Compton process emit synchrotron
photons in the IR--optical--UV range. 
As a consequence, the two emissions are expected to be strongly correlated.
Optical observations can therefore be crucial to understand the nature
of $\gamma$--ray emission in blazars since they allow to derive information
on both the seed photons and the radiating high energy electrons. 

Renewed interest in optical monitoring is motivated also by the
discovery that in the radio domain the flux of some blazars varies by
10--30\% on a timescale less than a day (Quirrenbach et al. 1992). 
The inferred brightness temperature, of the order of $10^{18}$ K, poses
severe difficulties to theoretical models, since it cannot be
reconciled with the Compton limit of  $10^{12}$ K for the commonly
estimated values of the beaming factor (see, e.g. Blandford, 1990).
Possible explanations are that these variations are due to regions with
extremely high beaming or to coherent processes or that they are
extrinsic, due to interstellar scintillation. 
Simultaneous optical and radio monitoring is thus critical to discriminate
different models.

In this paper we present the results of recent optical-UV observations
of the BL Lac object S5 0716+714. This source is an ideal target for both the 
research fields outlined above, since it has been detected several times 
in the $\gamma$-ray band and has shown intraday variability at optical 
and radio wavelengths (Wagner et al. 1996). 
The optical observations were carried out between Nov. 15, 1994  and Apr. 30, 
1995, the International Ultraviolet Explorer (IUE) pointings were 
on Feb. 27 and March 1, 1995 and cover the EGRET--CGRO viewing period 
(1995 Feb. 14--28).

The outline of the paper is as follows: in Sect. 2 we review the available 
information on S5 0716+714 at all wavelengths;
in Sects. 3 and 4 we describe and discuss the optical observations, taken
from different sites and telescopes, and the UV observation obtained
with IUE simultaneously with the phase 4 EGRET pointing;
in Sect. 5 we model the overall emission spectrum with a homogeneous
synchrotron self Compton (SSC) model and with a relativistic inhomogeneous jet
model. In Sect. 6 we draw our conclusions. 

\section {The BL Lacertae object S5 0716+714}

S5 0716+714 was discovered in the Bonn--NRAO radio survey (K\"uhr et al. 
1981) of flat spectrum radio sources with a 5 GHz flux greater than 1 Jy,
and was identified as a BL Lacertae object, because of the featureless
spectrum, by Biermann et al. (1981). 
From the stellar appearance on deep plates Stickel, Fried 
\& K\"uhr (1993) derived a lower limit of $z>0.2$ on the redshift, while
Schalinski et al. (1992) give $z>0.3$.
Proper motions of VLBI components have been observed (Schalinski et al.
1992), and would result in an apparent superluminal motion with
$\beta_{app}=4.6$ if $z=0.3$ and $H_0=50$ km s$^{-1}$ Mpc$^{-1}$. 

A high dynamic range VLA 1.5 GHz radio map shows a double structure 
(Antonucci et al. 1986) resembling that of an FR II radiogalaxy seen end--on,
but higher frequency (5 and 8.5 GHz) VLA maps do not show the
double structure (Wagner et al. 1996).
The FR II morphology conflicts with the suggestion that the parent
population of BL Lac objects should be FR I radiogalaxies (see e.g. Urry
\& Padovani 1995 for a review) but it is consistent with a more general
``unified" scheme whereby both FR I and FR II radiogalaxies have similar
relativistic jets in their cores (Maraschi \& Rovetti 1995).

A further ``anomaly" is that the observed ratio of the core to the
extended radio flux, which is $R=4.7$ at 5 GHz and $R\sim 1$ at 1.5 GHz
(Perley, Fomalont \& Johnston 1982), is at the low extreme of the
$R$--distribution of the radio--selected BL Lac objects (see e.g.
Ghisellini et al. 1993 and references therein). 

The radio core has an inverted spectrum [$\alpha=-0.42$ between 1.5 and 5
GHz; Perley, Fomalont \& Johnston 1980; $F(\nu)\propto \nu^{-\alpha}$]
The overall flat radio spectrum seems the result of the superposition
of the core and more extended and steeper components (see the
decomposition of the radio spectrum in Eckart et al. 1986), even if
the intraday radio variability suggests that at least some
of the radio emission may come from much smaller regions. 
                                                              
The source  was monitored in the mm band by Steppe et al. (1992, 1993) 
and Reich et al. (1993). 
At these frequencies (90--230 GHz) the flux varied by a factor 2--3 on
timescales of the order of a month. 
As shown by Reich et al. (1993), the radio flux at all the monitored
frequencies increased after Jan. 1992, the time of the first detection
by EGRET--CGRO.

In the far infrared it was detected by the Infrared Astronomical Satellite
(IRAS) at 12, 60, and 100 $\mu$m (Impey \& Neugebauer 1988).

Despite S5 0716+714 is a very interesting source it has been very poorly
observed in the past. Biermann et al. (1981) report the two photographic
magnitudes 13.2 and 15.5, the former derived from the POSS plate. The two 
previous visual magnitude estimates of 11 (K\"uhr et al. 1981) and $19\pm 1$ 
(Perley, Fomalont \& Johnston 1980) cannot be considered as fully 
reliable because no clear source for them is specified by these authors.
The 8--month long relative $R$ light curve, from September 1990 to 
March 1991 (Heidt \& Wagner 1996, Wagner et al. 1996), shows a 
variation amplitude of about 2.5 mag. No historic curve, based on archive 
plates, to search for possible long term flux changes, is available. 
Takalo, Sillanp\"a\"a~ \& Nilsson (1994) found a high level of optical 
polarization, which varied from 20\% to $\sim$ 16\% in two days.

In  X--rays the source was observed by HEAO1, EINSTEIN, 
and ROSAT (Biermann et al. 1981, 1992; Wagner, 1992;
Cappi et al. 1994; Comastri, Molendi \& Ghisellini 1995).
During the ROSAT observations the flux varied by 70\% in 1000 seconds
and by a factor 7 in 2 days (Wagner 1992, Cappi et al. 1994).
EGRET observed and detected S5 0716+714 several times from January
1992 (Lin et al. 1995; von Montigny et al. 1995).
The integrated flux above 100 MeV varied between 1.3$\pm$0.5 and
5.3$\pm$1.3$\times 10^{-7}$ photons cm$^{-2}$ s$^{-1}$.
Low statistics did not allow to derive a significant change of the
energy spectral index during variations of the flux 
(Lin et al. 1995). 
Combining observations of four different viewing periods, Lin et al. (1995)
found an average energy spectral index $\alpha_\gamma=0.85\pm 0.20$.

Following the discovery of intraday radio variability 
with a $\sim 10$\% amplitude (Heeschen et al. 1987), the source was 
monitored several times at radio wavelengths, always showing intraday 
variability (Quirrenbach et al. 1989, 1992). At optical frequencies,
variations on timescale of $\sim$1 day with amplitude of several tenths 
of magnitude are reported by Wagner et al. (1996).

A simultaneous optical ($R$ band) and radio (5 GHz) monitoring campaign,
performed for a month (February 1990), suggested that the  variations
in the two bands were correlated,
with amplitudes up to a factor 3 in the optical and 25\% 
in the radio, and with typical timescales of 1 and 7 days in both bands
(Wagner et al. 1990; Quirrenbach et al. 1991).
Further support to the suggestion of a correlated optical and radio
variability comes from the correlation between the radio spectral
index in the range 5--8.4 GHz and the optical flux (Wagner et al.
1996 and references therein). If the redshift $z>0.3$ 
the implied brightness temperature exceeds $10^{18}$ K 
(Quirrenbach et al. 1991).

\section{Observations and data analysis}

\subsection{Optical}
Photometry was carried out with several telescopes: the 1.05 m
astrometric telescope of the Observatory of Torino (B and R bands mainly), 
the 0.4 m automatic telescope of the University of Perugia (Tosti, Pascolini 
\& Fiorucci 1996), the 0.5 m telescope of the Astronomical Station of 
Vallinfreda, and the 0.4 m telescope of the IAS--CNR at the Observatory of 
Monte Porzio (the last two operated by the Rome group).
All telescopes were equipped with CCD cameras and filters $B$, $V$
(Johnson) and $R$, $I$ (Cousins). 
Data reduction was performed with different software packages: the
standard IRAF and MIDAS procedures and some others developed locally,
as the ROBIN program developed at the Observatory of Torino by L. 
Lanteri (private communication).
All images were bias and flat--field corrected. 
The comparison among data obtained in the same nights with different
telescopes yielded the same results within the errors, which are
typically 0.02 mag in $R$ and 0.03--0.05 mag in $B$. 

We did not find any published comparison sequence in the field
of S5 0716+714; only one nearby star was calibrated by 
Takalo, Sillanp\"a\"a~ \& Nilsson (1994).
We therefore calibrated some reference stars using several Landolt's (1992)
fields. The stars are indicated in the map of Fig. 1 and the corresponding
magnitudes are listed in Table 1; quoted errors include the uncertainty on
the absolute calibration, while magnitude differences between the reference
stars in a given filter are good within 0.02 magnitudes.
The star C of this sequence is the one
already measured by Takalo, Sillanp\"a\"a~ \& Nilsson (1994): our
magnitudes are consistent with theirs.

\begin{figure}
\epsfysize=3in
\epsffile{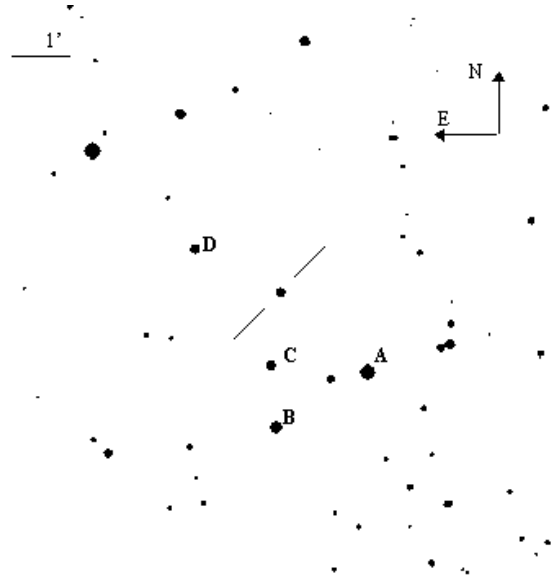}
\caption[]{
Finding chart of S5 0716+714. See Table 1 for the magnitudes
of the labeled reference stars.
}
\end{figure}

The reddening correction in the direction of S5 0716+714 was taken 
$A_V=0.23$. It was derived from the $N_{\rm H}$ value of 
$3.9\times 10^{20}$ cm$^{-2}$
(Ciliegi, Bassani \& Ca\-roli 1993), in agreement with the best fit values
of both UV and X--ray observations (Cappi et al. 1994), using the conversion 
formula by Shull \& van Steenberg (1985) and R=3.09 (Riecke \& Lebofsky 1985).
From the curve of Riecke \& Lebofsky (1985) we derived also $A_B=0.30$,
$A_R=0.19$, and $A_I=0.14$. In the following we use these values when
deriving fluxes from the apparent magnitudes. 

\begin{table}
\caption[]{Reference stars in the field of S5 0716+714}
\begin{tabular}{l l l l l}
\hline
      &$I$    &$R$ &$V$  &$B$     \\
\hline
 A &10.92$\pm$0.04  &11.21$\pm$0.04  &11.51$\pm$0.04  &12.07$\pm$0.06  \\
 B &11.79$\pm$0.05  &12.12$\pm$0.05  &12.48$\pm$0.05  &13.06$\pm$0.08  \\
 C &12.85$\pm$0.05  &13.18$\pm$0.05  &13.58$\pm$0.06  &14.17$\pm$0.08  \\
 D &12.97$\pm$0.04  &13.27$\pm$0.04  &13.66$\pm$0.04  &14.25$\pm$0.05  \\
\hline
\end{tabular}
\end{table}

The observation journal is reported in Table 2: here, for each date
are given the measured magnitudes and their uncertainties in the four 
filters together with the used telescope. In some occasions
several images were made in the course of the same night to search
for intraday variability; these nights are indicated by an asterisk.
The magnitude uncertainties have been 
evaluated by taking into account the mean dispersion of the internal error,
but not the errors on the absolute calibration.
The entire data set can be retrived via anonymous ftp at the address
$ftp.to.astro.it$ directory $/pub/blazars/$.

\begin{table*}
\begin{center}
\caption[]{Observed magnitudes of S5 0716+714}
\begin{tabular}{lllllll}
\hline
 Date      &  JD    &          $B$  &        $V$  &          $R_C$  &      $I_C$  & Tel.   \\
\hline 
14 Nov 1994 & 671.55 &             &                 & 13.52$\pm$0.03 &             & T  \\
        15  & 672.55 &             &                 & 13.73$\pm$0.02 &             & T  \\
        23  & 680.61 &             &                 & 13.86$\pm$0.02 &             & T  \\
        29  & 686.64 &             &                 & 13.65$\pm$0.02 &             & T  \\
11 Dec      & 698.50 &13.87$\pm$ 0.03&               &                &             & T  \\
02 Jan 1995 & 720.51 &14.25$\pm$ 0.03&               & 13.41$\pm$0.02 &             & T  \\
        03  & 721.40 &14.15$\pm$ 0.03&               & 13.30$\pm$0.02 &             & T  \\
        04  & 722.52 &14.09$\pm$ 0.03&               & 13.26$\pm$0.02 &             & T  \\
        06  & 724.39 &13.84$\pm$ 0.03&               & 13.04$\pm$0.02 &             & T  \\
      09 *  & 727.39 &13.70$\pm$ 0.03&               & 12.86$\pm$0.02 &             & T  \\
      14 *  & 732.47 &13.96$\pm$ 0.03&               &                &             & T  \\
        24  & 742.37 &13.80$\pm$ 0.03&               & 12.99$\pm$0.02 &             & T  \\
        25  & 743.33 &14.11$\pm$ 0.03&               & 13.25$\pm$0.02 &             & T  \\ 
        30  & 748.39 &13.77$\pm$ 0.03&               & 12.93$\pm$0.02 &             & T  \\ 
02 Feb      & 751.44 &13.77$\pm$ 0.03&               & 12.93$\pm$0.02 &             & T  \\
      04 *  & 753.47 &13.90$\pm$ 0.03&               & 13.05$\pm$0.02 &             & T  \\
        09  & 758.45 &13.75$\pm$ 0.03&               & 12.90$\pm$0.02 &             & T  \\
        14  & 763.35 &13.73$\pm$ 0.03&               & 12.88$\pm$0.02 &             & T  \\
        16  & 765.38 &13.68$\pm$ 0.03&               & 12.85$\pm$0.02 &             & T \\
        18  & 767.46 &13.64$\pm$ 0.03&13.22$\pm$ 0.02& 12.78$\pm$0.02 &             & T  \\
        19  & 768.42 &13.63$\pm$ 0.03&13.21$\pm$ 0.02& 12.78$\pm$0.02 &             & RT  \\
        20  & 769.25 &13.67$\pm$ 0.03&13.23$\pm$ 0.03& 12.80$\pm$0.02 &12.29$\pm$ 0.03& PR  \\
        22  & 771.37 &13.63$\pm$ 0.03&13.25$\pm$ 0.04& 12.83$\pm$0.02 &12.31$\pm$ 0.04& P  \\
      27 *  & 776.37 &13.88$\pm$ 0.03&13.49$\pm$ 0.02& 13.07$\pm$0.02 &             & PT  \\
      28 *  & 777.44 &13.80$\pm$ 0.03&13.42$\pm$ 0.02& 12.98$\pm$0.03 &12.45$\pm$ 0.03& R  \\
01 Mar      & 778.39 &13.94$\pm$ 0.03&               & 13.10$\pm$0.02 &             & T  \\
        02  & 779.37 &14.07$\pm$ 0.03&               & 13.24$\pm$0.02 &             & T  \\
        05  & 782.46 &14.29$\pm$ 0.03&13.83$\pm$ 0.02& 13.45$\pm$0.02 &12.98$\pm$ 0.05& R  \\
        06  & 783.30 &14.34$\pm$ 0.03&               & 13.51$\pm$0.02 &             & T  \\ 
        07  & 784.31 &14.30$\pm$ 0.03&13.96$\pm$ 0.03& 13.48$\pm$0.02 &12.96$\pm$ 0.03& P  \\ 
        09  & 786.41 &14.13$\pm$ 0.03&               & 13.32$\pm$0.02 &             & T  \\
        10  & 787.47 &14.17$\pm$ 0.03&               & 13.32$\pm$0.02 &             & T  \\
      11 *  & 788.39 &14.38$\pm$ 0.03&13.98$\pm$ 0.03& 13.53$\pm$0.02 &13.01$\pm$ 0.03& PT  \\
        14  & 791.31 &14.44$\pm$ 0.03&14.00$\pm$ 0.03& 13.59$\pm$0.02 &13.07$\pm$ 0.03& P  \\
        17  & 794.37 &               &14.41$\pm$ 0.03& 13.95$\pm$0.02 &13.40$\pm$ 0.03& P \\
        19  & 796.36 &14.81$\pm$ 0.03&14.36$\pm$ 0.03& 13.97$\pm$0.03 &13.44$\pm$ 0.05& R  \\
        20  & 797.26 &14.41$\pm$ 0.03&14.08$\pm$ 0.02& 13.64$\pm$0.02 &               & P  \\
            & 797.36 &14.34$\pm$ 0.03&13.97$\pm$ 0.03& 13.54$\pm$0.02 &13.03$\pm$ 0.03& R \\
        21  & 798.30 &14.57$\pm$ 0.03&14.15$\pm$ 0.03& 13.70$\pm$0.02 &             & T  \\
        22  & 799.37 &14.96$\pm$ 0.03&14.52$\pm$ 0.02& 14.07$\pm$0.02 &13.52$\pm$ 0.03& R  \\ 
        23  & 800.39 &15.06$\pm$ 0.03&14.64$\pm$ 0.03& 14.20$\pm$0.02 &13.63$\pm$ 0.03& P  \\ 
        24  & 801.37 &             &14.51$\pm$ 0.02& 14.05$\pm$0.02 &13.50$\pm$ 0.03& P  \\
        31  & 808.39 &14.34$\pm$ 0.03&13.97$\pm$ 0.03& 13.54$\pm$0.03 &12.98$\pm$ 0.04& P   \\
01 Apr      & 809.31 &14.47$\pm$ 0.03&14.05$\pm$ 0.02& 13.62$\pm$0.02 &13.14$\pm$ 0.03& R  \\
        03  & 811.40 &14.80$\pm$ 0.03&             & 13.85$\pm$0.02 &             & T  \\
        04  & 812.34 &14.71$\pm$ 0.03&             & 13.82$\pm$0.02 &             & T  \\
        05  & 813.31 &14.88$\pm$ 0.03&             & 13.96$\pm$0.02 &             & T  \\ 
        06  & 814.36 &14.77$\pm$ 0.03&             & 13.91$\pm$0.02 &             & T  \\ 
        07  & 815.33 &               &14.14$\pm$ 0.02& 13.70$\pm$0.02 &13.19$\pm$ 0.03& P  \\
        09  & 817.33 &14.63$\pm$ 0.03&14.25$\pm$ 0.02& 13.82$\pm$0.02 &13.28$\pm$ 0.03& P   \\
        12  & 820.35 &14.94$\pm$ 0.03&             & 14.05$\pm$0.02 &             & T  \\
        13  & 821.45 &14.83$\pm$ 0.03&             & 13.98$\pm$0.02 &             & T  \\
        14  & 822.38 &14.76$\pm$ 0.03&             & 13.94$\pm$0.02 &             & T  \\
        19  & 827.33 &             &14.22$\pm$ 0.03& 13.80$\pm$0.02 &13.29$\pm$ 0.03& P  \\
        30  & 838.38 &14.76$\pm$ 0.03&             & 13.95$\pm$0.02 &             & T  \\
\hline
\end{tabular}

The asterisk indicates the nights in which more than four observations with
the same telescope were performed.

Telescope identifications: R = Roma , P = Perugia, T = Torino
\end{center}
\end{table*}

The light curves in the four bands are shown in Fig. 2a,d; in Fig. 3 
the part of the R light curve (the most sampled in that period) corresponding 
to the EGRET pointing is shown in more detail.

\begin{figure}
\vskip 0.5 true cm
\epsfysize=3.7in
\epsffile{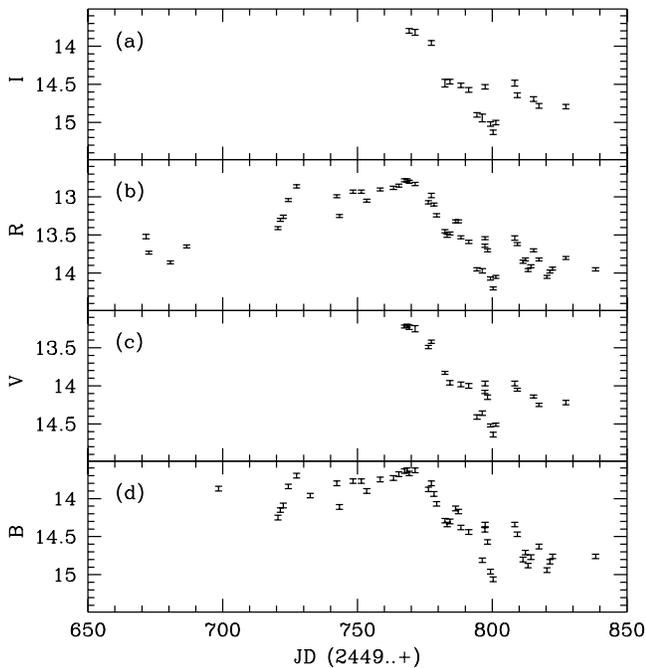}
\caption[]{Light curve of S5 0716+714 in the I, R, V, B bands.
}
\end{figure}

\begin{figure}
\vskip 1.5 true cm
\epsfysize=4.5in
\epsffile{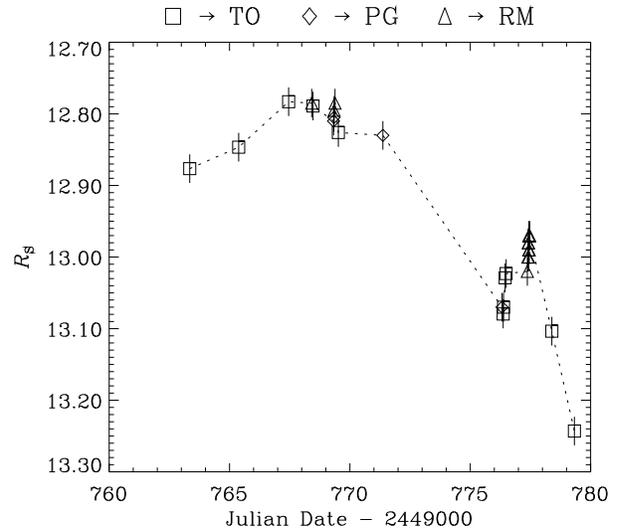}
\vskip -5 true cm
\caption[]{
Enlargement of the light curve in the $R$ filter
during the period of the EGRET pointing.
}
\end{figure}

\subsection{Optical and Ultraviolet spectra}

--- {\it Optical spectrum} ---
 
S5~0716+714 was observed with the 1.5 m Cassini telescope of the Loiano
Station, operated by the Observatory of Bologna. 
The 4000--8000 \AA~~ spectrum was obtained on March 19, 1995 at 00:58 UT,
with the BFOSC all--transmitting spectrograph. 
A Thomson $1024 \times 1024$-pixel coated CCD detector was used with 300
l/mm grating and 2".5 slit to give a spectral resolution of $\sim$ 25 \AA.  
The total integration time was 2400 s. 
The 2--dimensional frame was biased and flat--field corrected using
standard techniques, and for the extraction of the one--dimensional
spectrum the optimal extraction algorithm (Horne 1986) was adopted.
The wavelength--calibration was performed with a fifth--order
polynomial fitting to the dispersion curve of He--Ar comparison lamp
spectrum. 
The spectrum has been flux--calibrated with the observation of the
standard star BD~+33~2642 (Stone 1977) and corrected with the
local mean atmospheric extinction curve.
The zero point of the spectral--flux calibration is not very
accurate due to the poor seeing ($\sim$ 3 arcsec), and to the 
non-photometric conditions during the night.

The resulting spectrum is shown in Fig. 4, where no significant 
features are visible. 
The continuum spectrum was fitted with a power--law model assuming the
reddening correction value A$_V$ = 0.23  and the parametric
formula for the mean extinction law computed by Cardelli et al. (1989),
derived from the Riecke \& Lebofsky (1985) extinction curve. 
The best fit spectral index and associated 1$\sigma$ error is $\alpha$ =
0.95 $\pm$ 0.05. 

\vskip 0.5 true cm

\begin{figure}
\vskip -1.5 true cm
\epsfysize=3.5in
\epsffile{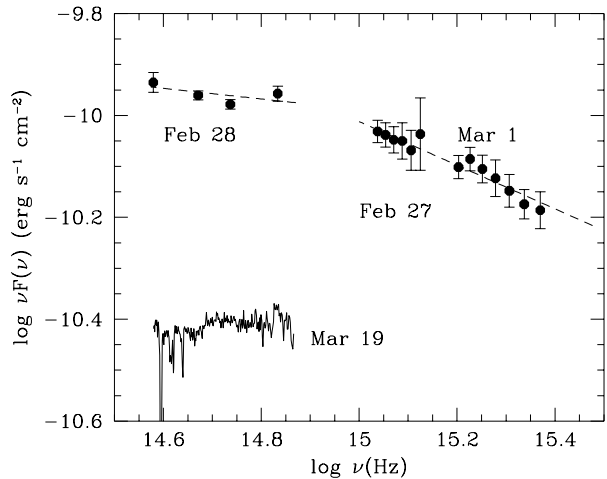}
\vskip -1 true cm
\caption[]{
The (binned) UV spectrum is plotted together with simultanoues optical data
and with the optical spectrum made 19 days after, as labelled.
}
\end{figure}

\noindent
--- {\it Ultraviolet spectrum} ---

S5 0716+714 was observed with the two cameras aboard IUE on February 27 
(LWP) and on March 1, 1995 (SWP, see Table 3). 
The spectral signal was extracted from the line-by-line image files
using the GEX routine (Urry \& Reichert 1988) running within MIDAS, and
flux calibrated using the curves provided by Bohlin et al. (1990, SWP)
and Cassatella et al. (1992, LWP); the resulting spectral flux
distribution, where no apparent emission or absorption features are
visible, is also given in Fig. 4. 
The flux was integrated and averaged over bins of 100 {\rm \AA}, and the
statistical uncertainties associated with the resulting flux points were
evaluated as in Falomo et al. (1993). To the flux errors a 1.25\% 
photometric error was added, according to Edelson et al. (1992). 
The SWP and LWP flux distributions were separately fitted to a power-law
model taking into account the reddening correction A$_V$ = 0.23, and the
extinction curve of Cardelli et al. (1989).
The fitted spectral parameters at the 1-$\sigma$ confidence level are
given in Table 3.

The optical photometry indicates that the source was stationary during the 2 
days which separate the IUE observations, thus, assuming absence of
significant interday UV variability a power--law fit has been attempted to the 
joined SWP and LWP spectral flux distributions (1230--2800 {\rm \AA}).
The resulting spectral index is intermediate between those obtained from the 
individual fits ($\alpha_{\nu}$ = 1.43 $\pm$ 0.03), with an acceptable value of 
$\chi^2_r$. This, together with the fact that the slope of the high energy 
part of the UV spectrum is not significantly larger than that of the lower 
energy region ($\Delta\alpha \sim 0.2$), 
excludes the measurable presence of a curvature or 
of a bump in the spectrum at $\sim$2000 {\rm \AA}. 

A power--law fit to the merged SWP and LWP spectra with the 
interstellar extinction taken as a free parameter yields a minimum 
$\chi^2$ for A$_V$ coinciding with the galactic value.
Including the simultaneous optical data (see the $B$, $V$, $R$ and $I$
data points in Fig. 4) the combined optical+UV fit yields
$\alpha$ =1.27$\pm$ 0.02, with $\chi^2_r=5.1$ (16 d.o.f.).
The BVRI data are fitted by a power law of slope $\alpha$ =1.1$\pm$ 0.1, 
significantly flatter than the UV slope, therefore one can infer 
that there is a break in the spectrum which occurs around the U band. 

We finally note that the flux at 1400 {\rm \AA} ($F_\nu$ = 3.21 $\pm$ 0.07 
mJy), 
is more than 20 times larger than that measured in 1980 (IUE exposure SWP 
9440; Pian \& Treves 1993).

\begin{table*}
\begin{center}
\caption[]{IUE observations}
\begin{tabular}{llllllll}
\hline
  &$\Delta\lambda$ &$\lambda$ (\AA) &$\alpha_\nu$  &$F(\nu)$ (mJy)  
&$\chi^2/n$ & d.o.f.\\
\hline
 SWP 54002 &1230-1930 &1600 &1.60$\pm$0.09 &3.98$\pm$0.07 &0.99 & 6  \\
 LWP 30122 &2200-2800 &2600 &1.39$\pm$0.23 &7.87$\pm$0.14 &0.25 & 5  \\
 SWP$+$LWP &1230-2800 &2000 &1.43$\pm$0.03 &5.45$\pm$0.08 &1.01 & 12 \\
\hline
\end{tabular}

$\Delta\lambda$ indicates the wavelength range used in the
analysis.

1 $\sigma$ errors are given both for the spectral index and for the flux.

The monochromatic flux $F(\nu)$ is calculated at the frequency corresponding to
the wavelength $\lambda$.
\end{center}
\end{table*}

\section{Results of the optical monitoring}
\subsection{The light curves}

From the light curves of Fig. 2a,d, one can see that the source brightened 
after JD=2449720, remained bright for $\sim$50 days, and then declined 
rapidly, to reach a mi\-nimum around JD=2449800. 
Note that in the `high' state the source almost reached its 
recorded historical maximum.

\begin{figure}
\epsfxsize=3.7in
\epsffile{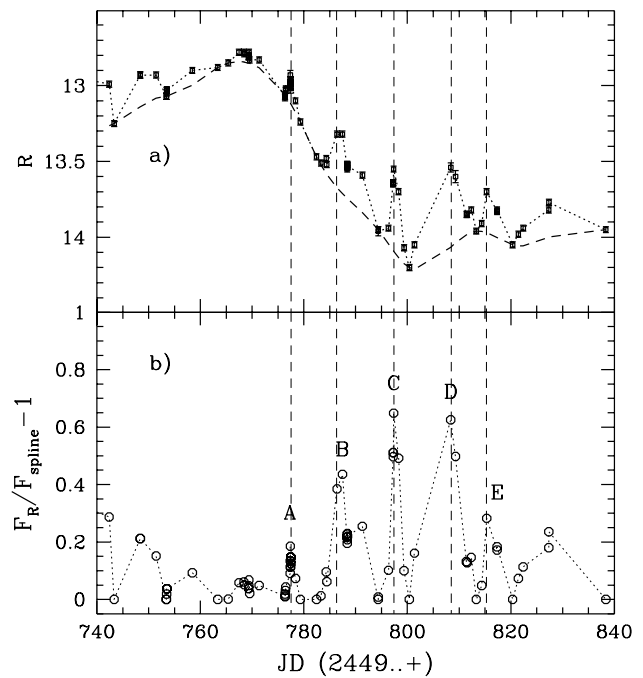}
\caption[]{
The light curve in the $R$ band (panel a) and the adopted spline function
passing through the local minima of the light curve (dashed line).
In panel b) we plot the ratio (minus 1) between the observed
flux and the spline value.
The labels and the vertical dashed lines identify interesting flaring events.
}
\end{figure}

Superposed to this relatively long term behaviour, there are fast 
fluctuations, whose amplitude appears wider in the `low state',
indicating that they could contain the same amount of energy
as in the `high state'.
To make this behaviour more apparent we modelled the `long' trend by 
interpolating a smooth curve through the local minima of the $R$ light
curve. We used a cubic spline interpolation, which is plotted as dashed line
in Fig. 5a. 
We stress that this spline interpolation 
crucially depends on the chosen minima of the light curve, and therefore it
should be considered only as a possible representation of the source 
behaviour.
The percent excesses between the observed fluxes and those derived
from this interpolating curve are shown in Fig. 5b: labels from A to E
correspond to five events of 'short' timescale. The event C is the best 
sampled, whereas the shape of the event B can depend on the adopted spline
and could be considered as the superposition of two (or even more) events.
Notice that the event A has the shortest duration and the smallest
variability amplitude.
For the best sampled peak in this curve, the rising and decaying 
times appear approximately equal, with no indications of a `plateau' longer
than one day.

\subsection{Intranight Variations}

Rapid variations of the flux of S5 0716+714 have been observed in 
different occasions (e.g. Wagner \& Witzel, 1995). 
In our campaign we also searched for intranight variations 
especially in the $R$ band (see Table 2).
Variations with an amplitude of about 5\% in few hours in 
the course of the same night were observed in some occasions.
Some of the resulting light curves in the $B$ and $R$ bands are shown 
in Figs. 6--7.
The lower panels in these figures show the difference between the
instrumental magnitudes of the two reference stars used in the night
(normalized to the mean value).

\begin{figure}
\vskip 0.5 true cm
\epsfxsize=3.3in
\epsffile{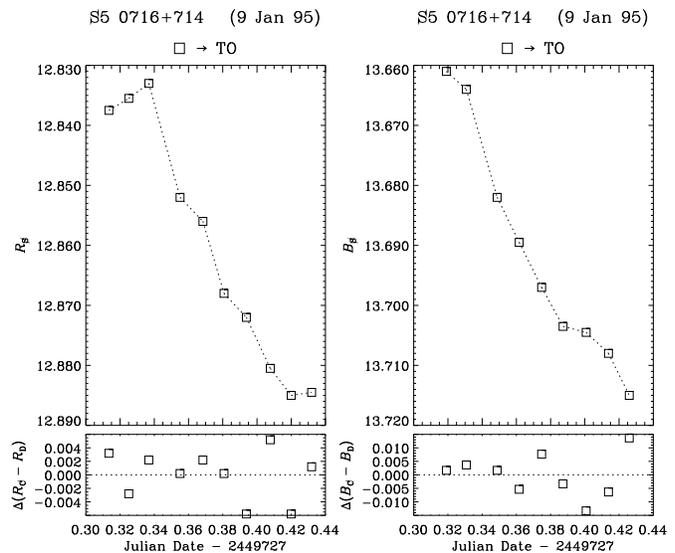}
\caption[]{
Intranight variations during the Jan. 9 1995 night, in the $R$ band
(left panel) and in the $B$ band (right panel)
}
\end{figure}

\begin{figure}
\vskip 0.5 true cm
\epsfxsize=3.3in
\epsffile{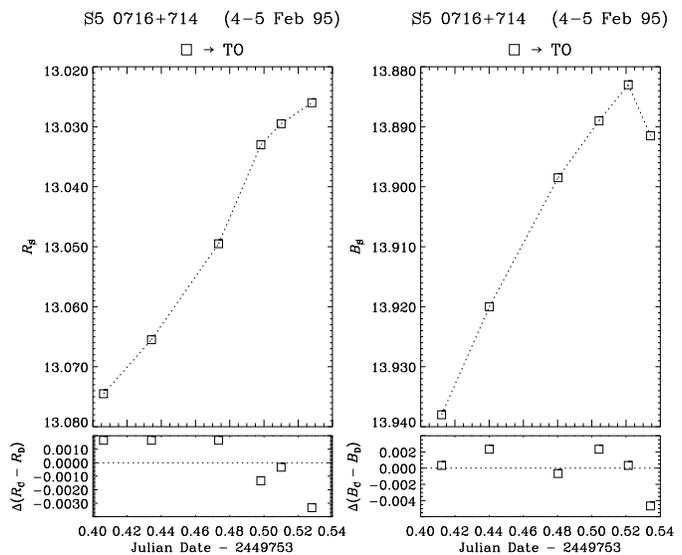}
\caption[]{
Intranight variations during the Feb 4 1995 night, in the $R$ band
(left panel) and in the $B$ band (right panel)
}
\end{figure}
 
 
Notice that in the two nights of 1995 Jan. 9 (Fig. 6) and Feb. 4 (Fig. 7) 
evident trends in the source brightness are apparent, like if our sampling 
was limited to a fraction of a longer time scale change. 
These cases, 
however, are important because the probability of their occurrence can be 
evaluated by means of the {\it run test} -- see, for instance, the 
introductory textbook by Barlow (1989) -- 
which is non--parametric and independent of the measured uncertainties. 
In the night of Jan 9, 1995, we have 9 data points in the $B$ band and 
other 10 in the $R$ band, both showing decreasing brightness trends and 
having each one two runs only. 
The expected run numbers for these two 
data samples are 5.44 ($B$ band) and 6.0 ($R$ band), corresponding to 2.5 
and 2.7 standard deviations, respectively. 
Their probabilities, estimated from a one--tailed gaussian distribution 
are $6.2\times 10^{-3}$ and $3.7\times 10^{-3}$; 
the combined chance probability to observe these trends at the same time 
and in the same direction in the two bands can be estimated as $1/2$ times 
their product, because the observations in the two bands can be considered 
independent, as indicated by the comparison star behaviour. 
We obtain then the value of $1.2\times 10^{-5}$, which gives 
a high statistical significance to this effect. 

Again on Feb 4--5 a well defined trend in the same 
direction is apparent in the two bands. 
The run test gives a total probability of $5.9\times 10^{-4}$, 
greater than the previous value because of 
the smaller number of data points, but statistically significant.

\subsection{Flux--spectral index correlation}

We computed the spectral index $\alpha$ (by a fit procedure), 
after dereddening the observed fluxes, for the nights in which 
at least three filter data are available, including intranight data. 
These nights cover the period from February 18 to the beginning of 
April in which the source brightness declined by 1.4 magnitudes. 
The resulting values of $\alpha$ vary between 0.81 and 1.15, with a mean 
value of 0.94, and do not have any significant correlation with the flux: 
the linear correlation coefficient with the V magnitude is only 0.15. 
In particular the spectral slopes measured when the source was at the 
maximum (February 19) and minimum (March 23) brightness were both equal 
to 0.99.

We also computed the 2--point spectral index $\alpha_{BR}$, which 
can be compared with the spectral index calculated using more filters.
The differences between these indices are usually less than the statistical
uncertainties and this gives confidence that $\alpha_{BR}$ traces the 
`true' optical spectral index. 
Again these values vary in the range 0.83--1.2.

\begin{figure}
\vskip -1 true cm
\epsfxsize=3.5in
\epsffile{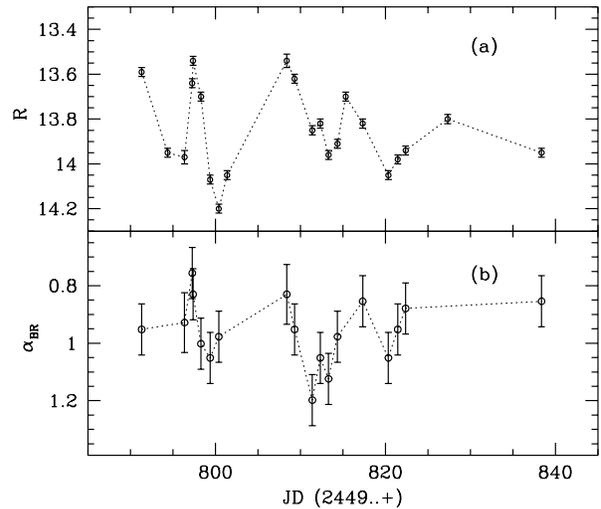}
\vskip -1 true cm
\caption[]{
The light curve in the $R$ band during the `low' state of
S5 0716+714 (panel a) compared with the corresponding spectral
index $\alpha_{BR}$ (panel b). }
\end{figure}

We found that when the source reached the ``low" state (after JD=2449790) 
the spectral index responded to the fast flux changes in the sense that 
the spectrum was flatter when the source was brighter. 
This is evident from Fig. 8 where $R$ and $\alpha_{BR}$ are plotted 
against time. 
The linear correlation 
coefficients of $\alpha_{BR}$ with $B$ and $R$ were 0.627 and 0.477,
respectively, and the corresponding chance probabilities are 
7.1$\times 10^{-3}$ and 5.3$\times 10^{-2}$. 
As pointed out by Massaro \& Trevese (1996) the latter estimate is 
biased towards a lower significance of the correlation and therefore 
this probability must considered as a conservative indication of the 
correct significance. 
A less biased estimate of the true correlation coefficient can be obtained
by taking the mean of the two previous values, and in this case the 
corresponding probability results equal to 2.1$\times10^{-2}$.
We can conclude that S5 0716+714 shows a significant ($\ge$ 95 \%) 
positive correlation between the spectral index and the flux only during
rapidly varying changes.

\section{Discussion}
\subsection{Short time variations}
The structure of the light curves of S5 0716+714 together with the color index
behaviour suggest that $two$ different processes are operating in the source:
the first one is responsible for the long term ``achromatic"
rise and decay of the brightness, while the second one is responsible
for the fast flares for which the color index ($\alpha_{BR}$)
is significantly bluer at higher intensity.
It is possible to interpret the first behaviour in terms of emission
produced in a large region, in which a stable energy injection mechanism
operates, at least on few month timescale.

The fact that the rising and decaying timescales of the short term flares are 
comparable, with much shorter plateaux, bears important consequences.
In fact, it is highly unlikely that these two timescales correspond
to the particle acceleration and cooling timescales, respectively, since
a priori there is no reason for them to be equal, and the observed equality 
would  be coincidental.
Instead, the equality is naturally explained in two frames:

1) As Wagner et al. (1993) suggested for the similar behaviour of
S4 0954+658, the characteristic variability pattern of the flares could
be caused by a curved trajectory of the relativistic emitting blob;
in this case the viewing angle $\theta$ (i.e.\ the angle between the blob 
velocity vector and the line of sight) varies with time and consequently the 
beaming or Doppler factor 
$\delta=[\Gamma(1-\beta \cos \theta)]^{-1}$
changes too. Since the observed monochoromatic flux can be assumed to be
proportional to $\delta^{3+\alpha}$, symmetric $\theta$ variations 
can account for the observed flare shape. In this case, for rather small
amplitude changes, one has

$$
{\Delta F \over F}\, \simeq \, 
(3+\alpha){\Delta\delta \over \delta}\, \simeq \, 
(3+\alpha)\beta\Gamma\delta \, \Delta\theta
\eqno(1)
$$
which shows that, for $\Gamma\sim\delta\sim 1/\sin\theta\sim 10$,
a flux variation of 20\% -- see the results of Sect. 5.2 -- 
implies a change of $\theta$ of less than 0.5 degrees only.

2) The rising and decay times are both associated with the light
crossing time.
This can happen if both the electron injection and cooling processes
operate on time-scales shorter than $R/c$ (where $R$ is the source radius), 
since in this case the light crossing time is the only characteristic 
time that can be observed.
On the other hand, the short duration of a plateau also requires that either 
the acceleration or the cooling times (or both) are not much less than $R/c$.
Assume in fact that $both$ these times are much shorter than $R/c$.
Then at any given time the received flux originates in a different
partial volume of the source.
Different times correspond to different sub-volumes of the source and, if
these regions are equal, the observed flare would have a plateau.

The situation is further complicated by the contribution of the steadier 
component, which dilutes the flux and spectral index variations.

We can envisage a test to discriminate between the two possibilities.
In case 1) the variable component should have a constant
spectral index: amplification due to different degree of beaming 
would in fact be `achromatic' (if there is no spectral break within the 
considered bands).
The sum of the variable (flatter) and the steady (steeper) components
results in a continuous flattening of the spectrum as the source
brightens. 
The flattest spectral index should correspond to the maximum flux.
In case 2) the variable component has instead a variable spectral index:
the maximum flux corresponds to integration over the source
volume, whose sub--regions contain electrons of different ages.
Therefore the maximum flux {\it does not} correspond to the flattest
spectral index.

Our data are not sufficient to discriminate between the two possibilities,
since a better sampling (in more bands, including IR and, possibly, UV) 
is needed.
Nevertheless, if possibility 2) turns out to be preferred by 
future investigations, it will be possible to draw important
consequences about the value of the magnetic field strength and the
particle energy.
We will derive these two parameters in the next section by independent 
arguments, and it is interesting to see if they agree.

Assume therefore that the rapid flares are due to quasi--istantaneous 
injection of electrons and rapid cooling.
In particular we require that the synchrotron and self Compton cooling times 
are equal to or shorter than $R/c$:

$$
t_{cool}\, = \, { 6\pi mc^2 \over \sigma_Tc\gamma_o B^2(1+U_S/U_B)} 
\, \le { R \over c},
\eqno(2)
$$
where $\sigma_T$ is the Thomson scattering cross section, $\gamma_o mc^2$ 
is the energy of the particles emitting in the optical, $U_B=B^2/(8\pi)$ 
is the magnetic energy density, and $U_S$ is the radiation energy density 
of the synchrotron emission (as measured in the comoving frame).
The observed frequency is given by
$$
\nu_o \, = \, {4\over 3} \gamma_o^2 \nu_B{\delta\over 1+z} ,
\eqno(3)
$$
where $\nu_B = eB/(2\pi mc)$ is the cyclotron frequency.
By relating the radius $R$ of the source with the observed variability 
timescale through $R\sim ct_{var}\delta/(1+z)$, we obtain the two limits:

$$
B\, \ge  \, 0.67 \, \nu_{15}^{-1/3} t_d^{-2/3}(1+U_S/U_B)^{-2/3}
\left({1+z \over \delta}\right)^{1/3} ,
\eqno(4)
$$
$$
\gamma_o\, \le \, 2\times 10^4 \nu_{15}^{2/3} 
\left[{ t_d (1+U_S/U_B)(1+z) \over \delta }\right]^{1/3} ,     
\eqno(5)
$$
where $B$ is in Gauss, $t_d$ is the variability timescale 
measured in days, and $\nu_o=10^{15}\nu_{15}$ Hz.
Note the relatively weak dependence on the beaming factor $\delta$.
Assuming $\delta/(1+z)=10$, $t_d=1$ day and $U_S/U_B=1$ (as indicated by
the ratio between the self--Compton and synchrotron luminosities), we find 
$B\ge 0.2$ G and $\gamma_o\le 10^4$.

\subsection{Homogeneous Synchrotron Self Compton Models}

In Fig. 9 we show the overall spectrum of S5 0716+714 from radio to 
$\gamma$--rays, calculated assuming a redshift $z=0.3$.
Solid symbols in the optical ($B$, $V$, and $R$) and UV bands correspond to 
observations simultaneous (Feb. 27 and Mar. 1, 1995) with those of the CGRO 
pointing, whose data are not yet public.
Empty squares are not simultaneous data taken from the literature.
Empty circles refer to simultaneous radio to UV data published by Wagner et 
al. (1996), the X--ray ROSAT spectrum is from Cappi et al. (1994) and
the $\gamma$--ray data are from Lin et al. (1995).

\begin{figure}
\vskip -2 true cm
\epsfysize=3.7in
\epsffile{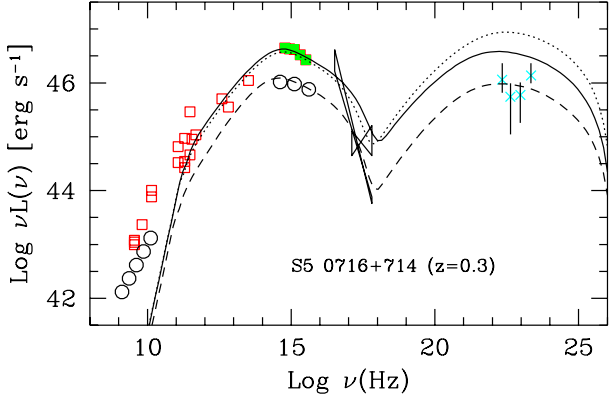}
\vskip -2 true cm
\caption[]{
Overall spectrum of S5 0716+714.
Empty circles are from Wagner et al. (1996), ROSAT X--ray data are
from Cappi et al. (1994) and $\gamma$--ray data are from Lin et al. (1995).
Filled squares are optical and UV data from our observations,
simultaneous (Feb 27 -- Mar 1, 1995) with EGRET observations (not public yet).
Empty squares are not simultaneous data from the literature.
Lines refer to homogeneous SSC models, as discussed in the text.
}
\end{figure}
     
The spectrum of S5 0716+714, like that of other 
$\gamma$--ray loud blazars, is characterized by two broad peaks
in $\nu$--$\nu F(\nu)$ plots, which can be interpreted as due to the maximum 
output of the synchrotron and Compton components.
Note that the knowledge of both the optical and UV 
spectra is crucial to define the frequency of the synchrotron peak.

Among the various ``Compton" models (see Introduction) we focus on
synchrotron self Compton (SSC) models, 
since no emission line has been observed in this source.
This limits the importance of photons produced externally to the jet or 
emitting blob, which could serve as seed photons to be Comptonized in the 
$\gamma$--ray band.

Consider a spherical blob, having the variability timescale radius 
$R=ct_{var}\delta/(1+z)$ (in the comoving frame), with a tangled magnetic 
field $B$, emitting synchrotron and self Compton radiation.
For appropriate particle distribution (to be better specified later),
the synchrotron and self Compton spectra peak at the observed frequencies
$\nu_S$ ($\sim \nu_o$), given by Eq. (3), and $\nu_C = (4/3)\gamma_o^2 \nu_S$. 
From the ratio $\nu_C/\nu_S$ we derive

$$
B\delta \, \simeq \, {\nu_S^2 \over 2.8 \times 10^6 \nu_C}(1+z)
\eqno(6)
$$

The ratio of the luminosities of the two peaks measures the relative
importance of the synchrotron and self--Compton powers.
Considering the Compton scattering process occurring entirely in the
Thomson regime, this ratio is directly related to the ratio between the
magnetic and radiation energy densities of the synchrotron photons: 

$$
{L_S \over L_C} \, \simeq \, {U_B \over U_S} \, =
\, {4\pi R^2cU_B \delta^4 \over L_S}
\eqno(7)
$$
Then
$$
B\delta^3 \, \simeq \, (1+z)\left( { 2L_S^2 \over L_C c^3 t^2_{var} }
\right)^{1/2}
\eqno(8)
$$
Equations (6) and (8) can be solved, yielding

$$
\delta\, \simeq \, 1.67\times 10^4 \left( {\nu_C\over t_d\nu_S^2} \right)^{1/2}
\, \left({ L^2_{S,45} \over L_{C,45} } \right)^{1/4}
\eqno(9)
$$
$$
B\, \simeq \, 2.14\times10^{-11} (1+z){ \nu_S^3 t_d^{1/2} \over \nu_C^{3/2} }
\left( { L_{C,45} \over L^2_{S,45} }\right)^{1/4}
\eqno(10)
$$
where $L_{S,45}$ and $L_{C,45}$ are measured in units of $10^{45}$ erg
s$^{-1}$ and $t_d$ is $t_{var}$ in days. 
Despite their simplicity, Eqs.\ (9) and (10) are a very powerful tool
for deriving $B$ and $\delta$, but we must stress that for a reliable
estimate we need to determine with great accuracy the simultaneous values
of $\nu_C$ and $\nu_S$, together with an estimate of the variability
timescales. 

For S5 0716+714, in the high state 
we have $L_S=2.2\times 10^{47}$ erg s$^{-1}$, 
$L_C=4.4\times 10^{47}$ erg s$^{-1}$,
$\nu_S=10^{15}$ Hz, $\nu_C=8\times 10^{22}$ Hz, and $t_{var}=2$ days.
With these values we derive $B\simeq 0.5$ Gauss, $\delta\simeq 11$, and
$\gamma_o\simeq 7.7\times10^3$.
Notice that these $B$ and $\gamma_o$ values are in agreement with the 
limits obtained in Sect. 5.1. 

To calculate the emission spectrum we need to know the energy distribution
of the emitting electrons. 
In a simple model with continuous injection of $Q(\gamma)\propto
\gamma^{-s}$ [cm$^{-3}$ s$^{-1}$] relativistic electrons 
throughout the source, the relativistic electron density $N(\gamma)$ can 
be obtained by solving the well known steady continuity equation, leaving
as free parameters the injected power $L_{inj}$, the spectral index $s$ 
of the injected distribution and 
the extremes of the energy range $\gamma_{min}$ and $\gamma_{max}$ 
of the injected electrons (see Ghisellini 1989 for details). 
For $s=2.6$, the resulting $N(\gamma)$ can be approximated with a double power 
law, with indices 2 and 3.6 below and above $\gamma_{min}$, respectively.
Therefore we can identify $\gamma_{min}$ as $\gamma_o$.
The emission spectra, computed for low (dashed line) and high (solid and
dotted lines) luminosity states are shown in Fig. 9. 
The two states correspond to an increase of the injected power 
$L_{inj}$ from 10$^{43}$ erg s$^{-1}$ to $3.3\times 10^{43}$ erg s$^{-1}$.      
The values of the other model parameters were $R=3\times 10^{16}$ cm,
$\delta = 12$, $\gamma_{min} = 3 \times 10^3$ (or $4\times 10^3$, dotted line), 
$\gamma_{max}= 10^5$, and $B=0.48$ (or 0.84, solid line) G.
The two high state spectra differ for the values of $B$ and $\gamma_{min}$:
the one represented by the solid line has an increased magnetic field 
to maintain the ratio $U_S/U_B$ constant and close to unity, while in 
the other the magnetic field is assumed to remain constant, and consequently 
the ratio $U_S/U_B$ increases.

These two choices, both reasonable from a physical point of view,
yield  $\gamma$--ray luminosities which are a factor of 3 to 10
greater than the EGRET luminosities of 1992--1993.
Furthermore, the emission is expected to vary
simultaneously at all frequencies from the mm band up to the GeV
band, although the amplitude of the synchrotron variations can be
smaller than the Compton one.

\subsection{Inhomogeneous jet model}

We adopt the model of Ghisellini \& Maraschi (1989) (see also
Ghisellini, Maraschi \& Treves 1985; Maraschi, Ghisellini \&
Celotti 1992) in order to check if relaxing the hypothesis of
homogeneity changes substantially the inferred physical parameters.
In this jet model the density of the emitting particles and the magnetic
field strength decrease along the jet as power law functions of the distance 
$x$ from the jet apex. 
Relativistic particle are assumed to have the same power law distribution,
all along the jet, but the high energy cut-off (and the maximum
emitted frequencies) are function of $x$. 
Spectral break and different spectral indices in the radiation spectrum
are due to the different superpositions of the local spectra.
The inner part of the jet has a parabolic shape, with cross sectional radius
$r \propto x^{1/2}$, while the outer part is conical ($r \propto x$). 
In addition, the emitting plasma flowing in the jet is assumed to
accelerate outwards in the parabolic region, while it has a constant
bulk velocity in the conical part. We also fixed the dependence
of the magnetic field with the cross sectional radius $r$ of the jet,
setting it to be $B\propto r^{-1}$, and the shape of the particle distribution
($N(\gamma)\propto \gamma^{-2}$, resulting in a fixed local spectral 
index, $\alpha_o=0.5$).
We furthermore required that the photon--photon opacity must not be 
important all along the jet and that the region producing most of the 
optical can vary in a $\sim$1 day timescale.
With all these constraints, the set of parameters which better interpolate
the data is rather well defined.
These are listed in Table 4, and correspond to the spectra shown in Fig. 10. 
We also plot, for the low state, the emission coming from different
regions of the jet. 

\begin{figure}
\vskip -2 true cm
\epsfysize=3.8in
\epsffile{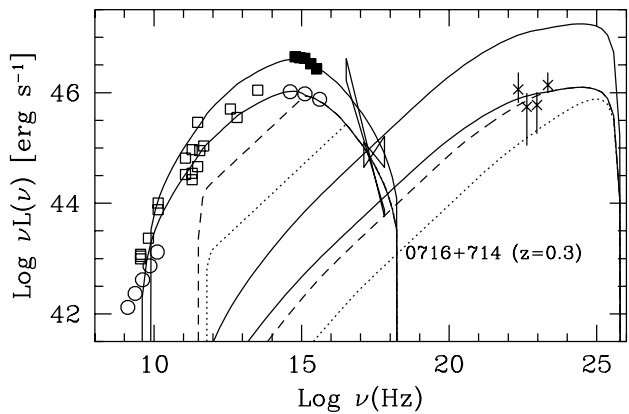}
\vskip -2 true cm
\caption[]{
The overall spectrum of S5 0716+714 interpreted as
emission from an inhomogeneous relativistic jet.
The solid lines refer to the emission from the entire jet, in the `high' 
and `low' state, according to the parameters listed in Table 4.
The synchrotron and the self Compton components are shown separately.
Dashed and dotted lines refer to the emission from
regions within $x=500 x_0$ and
within $x=50 x_0$ from the jet apex, respectively. See Table 4 for the
definition of $x_0$.
}
\end{figure}

The bulk of the $\gamma$--ray emission (up to a few GeV) is produced where the
parabolical and the conical part of the jet connect, at distance $x\sim
5\times 10^{17}$ cm from the jet apex, where the cross sectional radius
of the jet is $r\sim 2\times 10^{16}$ cm.
As explained in Ghisellini, Maraschi \& Treves (1985) and in Ghisellini
\& Maraschi (1989), this is expected in this class of models, due to
the dependence of the Compton flux as a function of distance $x$,
which is different in the parabolic and the cone zones.
However, as can be seen also in Fig. 10, higher frequencies
(both in the synchrotron and the self Compton branches) are
produced closer in.
This is not the consequence of the particular set of parameters,
but is again a general characteristic of the model: requiring to
fit the spectral change in the IR--UV--soft X--ray spectrum
as a superposition of local spectra (each with a different high energy
cut--off), inevitably implies higher frequencies be produced
in the innermost regions of the jet.
This characteristic can well be tested with future $\gamma$--ray
satellite, such as GLAST, able to observe up to the 100 GeV band,
where the emission is predicted to vary more dramatically and with
the shortest timescales.
Less dramatic differential variability is however expected also
in the different EGRET energy bins.
In the intermediate region between the paraboloid and the cone
the magnetic field strength is $B\sim 0.5$--1 G, the bulk 
Lorentz factor is $\Gamma\simeq 12$, the corresponding beaming factor is
$\delta\simeq 14$, and the magnetic and radiation energy densities are 
close to equipartition.

In conclusion, we derive a set of parameters which agree well with the
findings of the homogeneous model. 
With two important differences: 

1) In the inhomogeneous jet model the synchrotron X--ray emission 
comes from the inner part of the paraboloid, at distances 
$x\sim 10^{15}-10^{16}$ cm from the jet apex (and cross sectional 
radii $r\sim 10^{15}-3.5\times 10^{15}$ cm. 
Correspondingly, the X--rays should vary with the minimum 
timescale (the observed $t_{var}\sim 1$ hour), possibly explaining the
very fast variability reported by Cappi et al. (1994). 

2) In the inhomogeneous jet model variability is simultaneous only 
in selected bands. 
This can be seen in Fig. 10 where the dashed and dotted lines
correspond to the synchrotron and the self--Compton emission
arising in same regions of the jet: $\sim 1$ GeV emission should
vary fast and simultaneously with 1--10 keV X--rays, while
$\sim 100$ MeV $\gamma$--rays should vary simultaneously with the optical--UV
emission and be characterized by a somewhat longer timescale.
Spectral index variations in the IR--UV are possible if a perturbations
travelling along the jet activates different regions, starting
from the innermost one, responsible for the X--rays (for a detailed
discussion, see Celotti, Maraschi \& Treves, 1991).
In this case the higher frequencies vary first, followed by the lower ones.
Note that the predicted correlation between the flux and
the spectral index is in the same sense as observed: flatter
slopes when the flux is brighter.

Finally, note that in the inhomogeneous jet model, as well as
in the homogeneous one, one cannot explain simultaneous intraday
variability: even in the favorable case of a perturbance travelling
down the jet, close to the line of sight, one should observe a
time delay between the optical and the radio emission
(even if shortened by relativistic effects).

\begin{table*}
\begin{center}
\begin{tabular}{l l l l l l}
\hline
  &Low state    &Low state      &High state  &High state  &Note    \\
  &Paraboloid   &Cone           &Paraboloid  &Cone        &  \\
\hline
&&&&&\\
 $x_0$ (cm)        &$10^{15}$         &$5\times 10^{17}$   
                    &$10^{15}$         &$5\times 10^{17}$ &initial distance \\
 $x_{max}$ (cm)    &$5\times 10^{17}$ &$10^{19}$   
                    &$5\times 10^{17}$ &$10^{19}$         &final distance \\
 $B_0$ (Gauss)     &12                &0.54                
                    &12                &0.54              &initial magnetic field    \\
 $N_0$             &$3\times 10^5$    &$2.8\times 10^3$   
                    &$6.7\times 10^5$  &$6.3\times 10^3$  &initial electron density \\
 $\alpha_0$        &0.5               &0.5               
                    &0.5               &0.5               &local spectral index  \\
 $m$               &1                 &1                 
                    &1                 &1                 &$B\propto r^{-m}$ \\
 $n$               &1.5               &1.5               
                    &1.5               &1.5               &$N\propto r^{-n}$ \\
 $a$               &0.4               &0               
                    &0.4               &0                 &$\Gamma\propto x^a$    \\
 $\gamma_{max}$    &$4.2\times 10^5$  &$1.4\times 10^4$            
                    &$4.2\times 10^5$  &$1.4\times 10^4$  &                       \\
 $\Gamma_0$        &1                 &12                
                    &1                 &12                &initial bulk Lorentz factor \\
 $\theta$ (degree) &4                 &4                
                    &4                 &4                 &viewing angle    \\
\hline
\end{tabular}
\end{center}
\caption[]{Input parameters for the inhomogeneous jet model.
Note: $z=0.3$ is assumed. For the parabolic part, the cross sectional
radius $r$ scales with distance $x$ as $r\propto x^{1/2}$.}
\end{table*}

\section{Conclusions}

Dense optical monitoring is a very powerful tool to better understand the
radiation processes operating not only in the optical band, but also
in the high energy part of the spectrum, and to put models for
the  $\gamma$--ray emission on test.

By themselves, the optical data we have obtained for S5 0716+714
are indicative of a complex behaviour 
possibly characterized by more than one component.
The dependence of the spectral index on the fast variations of the flux and 
its insensitivity to longer-term trends, suggest localized injections 
of energy in the form of fresh and rapidly cooling relativistic particles, 
superimposed to a steadier emission, possibly coming from larger regions.
Other constraints can be obtained by the shape of the fast flares, which
seem symmetric and without plateau. 
In particular, this allows to derive a lower limit on the magnetic field 
strength.
It is very interesting to verify if other sources behave like
S5 0716+714 in this respect, to assess if this is a fundamental feature
to be explained by emission models for blazars in general. 
Indeed, some of the best monitored surces (e.g. 3C 66A, Takalo et al. 1996;
OJ 287, Sillamp\"a\"a~ et al. 1996; 0954+658, Wagner et al. 1993;
PKS 2155--304, Pian et al. 1997; PKS 0422+004, Massaro et al. 1996),
seem to confirm this behaviour.

S5 0716+714 was very bright in the optical during the period of our 
observations, and in particular during the period of the EGRET pointing.
Since electrons radiating optical synchrotron photons would emit 
$\gamma$--rays by the inverse Compton process, one inevitably predicts 
that the source  had to be bright also in the EGRET band. 
The exact amount of the expected $\gamma$--ray emission is of course
model--dependent (this is the reason why simultaneous observations
are so important), but we can restrict the existing possibilities
by noting that the absence of emission lines in the S5 0716+714 spectrum 
makes inverse Compton off photons produced $externally$ to the jet
less likely.
For S5 0716+714, a simple homogeneous and one--zone model is appropriate
to describe the optical/$\gamma$--ray behaviour, because the 
regions emitting both these frequency bands are cospatial. 
This remains true in more realistic inhomogeneous models, which can
better explain the entire electromagnetic output of the source,
including the radio, far infrared, and X--ray bands. 
These models can be put on test by simultaneous monitoring
in the above mentioned bands.
In fact, a key ingredient of the inhomogenous
jet model we used is that smaller jet regions,
located at short distances from the jet apex, are responsible for the 
high energy steep tail of the X--ray synchrotron spectrum, while
larger and outer regions mainly contribute to the radio--to--IR band by 
synchrotron, and to hard X--rays by self-Compton emission.
Perturbances travelling along the jet could be mapped in the frequency
space, with shorter variability timescales corresponding to higher
synchrotron and inverse Compton frequencies.
In addition, there should be time delays in the light curves at different
frequencies.

If the seed photons to be upscattered in energy are local synchrotron
photons, then we predict that S5 0716+714 should have reached, during 
the February 1995 EGRET observation, its brightest ever detected 
$\gamma$--ray state, with a flux a factor 3--10 higher than in
the 1992-1993 EGRET observing periods.
Furthermore, the simple SSC models presented here assume that the bulk of 
the optical and $\gamma$--ray luminosities are produced in the same region,
and therefore the flux in both bands should vary together, without delay.
In particular, we observed a decline of the optical flux on a time scale of 
a few days, and we expect that a similar trend should also be found in the 
$\gamma$-ray band if the statistical quality of the data is good enough.
If the $\gamma$--ray emission is not linked with the decline of the 
bulk of the optical emission, one is still left with the interesting
alternative that the $\gamma$--rays are connected with the optical fast flares.
On the other hand, the absence of any optical--$\gamma$--ray correlation 
would be a severe problem for all simple SSC models.


\begin{thebibliography}{}

\bibitem[]{}

\bibitem[]{}
Antonucci, R.R.J., Hickson, P., Olszewski, E.W. \& Miller, J.S.,
1986, AJ 92, 1.

\bibitem[]{}
Barlow, 1989, Statistics, J. Wiley \& Sons, Chapter 8 

\bibitem[]{}
Biermann, P.L., et al. 1981, ApJ 247, L53

\bibitem[]{}
Biermann, P.L., Schaaf, R., Pietsch, W., Schmutzler, T., Witzel, A.
\& K\"uhr, H., 1992, A\&AS 96, 339.

\bibitem[]{}
Blandford, R.D., 1990, in Active Galactic Nuclei, eds. T.J.L. Courvoisier, \&
M. Mayor (Berlin)

\bibitem[]{}
Blandford, R.D. \& Levinson, A., 1995, ApJ 441, 79

\bibitem[]{}
Blandford, R.D., 1993, in Compton Gamma Ray observatory, ed. N. Gehrels
\& M. Friedlander (New York: American Institute of Physics), 533

\bibitem[]{}
Bohlin, R., Harris, A.W., Holm, A.V., \& Gry, C., 1990, ApJS 73, 413

\bibitem[]{}
Cappi, M., Comastri, A., Molendi, S., Palumbo, G.G.C., Della Ceca, R. \&
Maccacaro, T., 1994, MNRAS 271, 438

\bibitem[]{}
Cardelli, J.A., Clayton, G.C., Mathis, J.S. 1989, ApJ 345, 245.

\bibitem[]{}
Cassatella, A., Gonzalez-Riestra, R., Imhoff, C., Oliversen, N., \& 
Lloyd, C., 1992, A\&A 256, 309


\bibitem[]{}
Celotti, A., Maraschi, L. \& Treves, A., 1991, ApJ, 377, 403

\bibitem[]{}
Ciliegi, P., Bassani, L. \& Caroli, E., 1993, ApJSS 85, 111

\bibitem[]{}
Comastri, A., Molendi, S. \& Ghisellini, G., 1995, MNRAS 277, 297

\bibitem[]{}
Dermer, C. \& Schlickeiser, R., 1993, ApJ 416, 458

\bibitem[]{}
Dondi, L. \& Ghisellini, G., 1995, MNRAS 273, 583


\bibitem[]{}
Eckart, A., Witzel, A., Biermann, P., Johnston, K.J., Simon, R., Schalinski,
C. \& K\"uhr, H., 1986, A\&A 168, 17

\bibitem[]{}
Edelson, R., Pike, G. F., Saken, J. M., Kinney, A., \& Shull, J. M.,
1992 ApJS, 83, 1

\bibitem[]{}
Fichtel, C.E. 1994, ApJS 90, 917

\bibitem[]{}
Falomo, R., Treves, A., Chiappetti, L., Maraschi, L., Pian, E., \& 
Tanzi, E. G. 1993, ApJ 402, 532

\bibitem[]{}
Ghisellini, G., 1989, MNRAS 238, 449

\bibitem[]{}
Ghisellini, G., Maraschi, L. \& Treves, A., 1985, A\&A 146, 204

\bibitem[]{}
Ghisellini, G. \& Maraschi, L., 1989, ApJ 340, 181

\bibitem[]{}
Ghisellini, G, Padovani, P, Celotti, A. \& Maraschi, L., 1993, ApJ
407, 65

\bibitem[]{}
Ghisellini, G. \& Madau, P., 1996, MNRAS 280, 67

\bibitem[]{}
Hartman, R.C., Webb, J.R., Marscher, A.P., et al., 1996, ApJ 461, 698

\bibitem[]{}
Heeschen, D.S., Krichbaum, T.P., Schalinski, C.J. \& Witzel, A.,
1987, AJ 94, 1493


\bibitem[]{}
Heidt, J. \& Wagner, S.J., 1996, A\&A 305, 42

\bibitem[]{}
Horne, K., 1986, PASP 98, 609

\bibitem[]{}
Impey, C.D. \& Neugebauer, G., 1988, AJ 95, 307

\bibitem[]{}
K\"uhr, H., Pauliny--Toth, I.I.K., Witzel, A. \& Schmidt, J., 1981,
AJ 86, 854

\bibitem[]{}
K\"uhr, H., Witzel, A., Pauliny--Toth, I.I.K. \& Nauber, U., 
1981, A\&AS 45, 367 


\bibitem[]{}
Landolt, A.U., 1992, AJ 104, 340


\bibitem[]{}
Lin, Y.C., Bertsch, D.L., Dingus, B.L., et al., 1995, ApJ 442, 96

\bibitem[]{}
Maraschi, L, Ghisellini, G. \& Celotti, A., 1992, ApJ 397, L5

\bibitem[]{}
Maraschi, L. \& Rovetti, F., 1995, ApJ 436, 79 


\bibitem[]{}
Massaro, E., Nesci, R., Maesano, M., et al., 1996, A\&A 314, 87

\bibitem[]{}
Massaro, E. \& Trevese, D., 1996, A\&A 312, 810

\bibitem[]{}
McBreen, B., 1979, A\& A 71, L19

\bibitem[]{}
Perley, R.A., Fomalont, E.B. \& Johnston, K.J., 1980, AJ 85, 649

\bibitem[]{}
Perley, R.A., Fomalont, E.B. \& Johnston, K.J., 1982, ApJ 255, L93

\bibitem[]{}
Pian, E. \& Treves, A., 1993, ApJ 416, 130

\bibitem[]{}
Pian, E., Urry, M.C., Treves, A., et al., 1997, ApJ, in press.

\bibitem[]{}
Quirrenbach, A., Witzel, A., Krichbaum, T., Hummel, C.A., Alberdi, A.
\& Shalinski, C., 1989, Nat 337, 442

\bibitem[]{}
Quirrenbach, A., Witzel, A., Wagner, S., et al., 1991, ApJ 372, L71

\bibitem[]{}
Quirrenbach, A., Witzel, A., Krichbaum, T.P., et al., 1992, A\&A 258, 279

\bibitem[]{}
Reich, W., Steppe, H., Schlickeiser, R., Reich, P., Pohl, M., Reuter, H.P., 
Kanbach, G \& Sch\"onfelder, V., 1993, A\&A 273, 65

\bibitem[]{}
Riecke, G.H. \& Lebofski, M.J., 1985, ApJ 288, 618

\bibitem[]{}
Schalinski, C.J., Witzel, A., Kircbaum, T.P., Hummel, C.A., Quirrenbach, 
A. \& Johnstone, K.J., 1992, in Variability in Blazars, ed. E. Valtaoja
\& M. Valtonen, (Cambridge Univ. Perss), p.221



Shull, J.M. \& Van Steenberg, M.E., 1985, ApJ 294, 599

\bibitem[]{}
Sikora, M., Begelman, \& Rees, M.J., 1994, ApJ 421, 153

\bibitem[]{}
Sillanp\"a\"a, A., Takalo, L.O., Lehto, H.J., et al., 1996, A\& A, 305, L17



\bibitem[]{}
Steppe, H., Liechti, S., Mauersberg, R., K\"ompe, C., Brunswig, W. \&
Ruiz--Moreno, M., 1992, A\&AS 96, 441

\bibitem[]{}
Steppe, H., Paubert, G., Sievers, A., et al., 1993, A\&AS 102, 611 

\bibitem[]{}
Stickel, M., Fried, J.W. \& K\"uhr, H., 1993, A\&AS 98, 393

\bibitem[]{}
Stickel, M., Padovani, P., Urry, C.M., Fried, J.W. \& K\"uhr, H.,
1991, ApJ 374, 431

\bibitem[]{}
Stone, R.P.S., 1977, ApJ 218, 767

\bibitem[]{}
Takalo, L.O., Sillanp\"a\"a~ \& Nilsson, K., 1994, A\&AS 107, 497

\bibitem[]{}
Takalo, L.O., Sillanp\"a\"a, A. Letho, H.J., et al., 1996, A\& ASS, 120, 313

\bibitem[]{}
Thompson, D.J., Bertsch, D.L., Esposito, J.A., et al., 
1995, ApJS 101, 259

\bibitem[]{}
Tosti, G., Pascolini,  \& Fiorucci, M., 1996, PASP 108, 706

\bibitem[]{}
Urry, C.M. \& Padovani, P., 1995, PASP 107, 803

\bibitem[]{}
Urry, C.M., \& Reichert, G. 1988, IUE NASA Newsletter, 34, 96


\bibitem[]{}
von Montigny, C., Bertsch, D.L., Chiang, J., et al., 1995, ApJ 440, 525

\bibitem[]{}
Wagner, S.J., Sanchez--Pons, F., Quirrenbach, A. \& Witzel, A.,
1990, A\&A 235, L1

\bibitem[]{}
Wagner, S.J., 1992, in ``X--ray emission form AGN and CRB", MPE proceedings,
p. 97

\bibitem[]{}
Wagner, S.J., Witzel, A., Krichbaum, T.P. et al.,
1993, A\&A 271, 344


\bibitem[]{}
Wagner, S.J. \& Witzel, A., 1995, ARA\&A 33, 163

\bibitem[]{}
Wagner, S.J., Camenzind, M., Dreissigacker, O., et al., 1995, A\&A 298, 688
 
\bibitem[]{}
Wagner, S.J., Witzel, A., Heidt, J., et al., 1996, AJ 111, 2187
  

\end{thebibliography}
\end{document}